# Strain controllable magnetocrystalline anisotropy in FeRh/MgO bilayers

Henry Hoffmann[1,*,] Eun Sung Jekal [1]

1 Department of Materials, ETH Zurich, Zurich 8093, Switzerland

* Correspondence: henry.hoffmann@mat.ethz.ch, Tel.: +41 76 470 62 46

**Abstract**

Ultra-thin film of FeRh on insulator MgO substrate has been investigated using ab-initio electronic structure calculations. From this calculation, we have found the interesting effect of epitaxial strain on the magnetocrystalline anisotropy (MCA). Analysis of the energy and k-resolved distribution of the orbital character of the band structure reveals that MCA largely arises from the spin-orbit coupling (SOC) between $d_{x^2-y^2}$ and d $xz,yz$ orbitals of Fe atoms at the FeRh/MgO interface. We demonstrate that the strain has significant effects on the MCA: It not only affects the value of the MCA but also induces a switching of the magnetic easy axis from perpendicular to in-plane direction. The mechanism is the strain-induced shifts of the SOC d-states. Our work demonstrates that strain engineering can open a viable pathway towards tailoring magnetic properties for antiferromagetic spintronic applications.

## 1. Introduction

Nowadays, the field of antiferromagnetic (AFM) spintronics has been treated as an promising material in the science community [1-10]. First priority reason is that they do not produce stray fields when it is used to device such as random access memory (RAM) [1-7]. Because structures being similar to their ferromagnetic (FM) counterparts, AFM spintronics complements or replaces ferromagnets by antiferromagnets in the active components of spintronic devices [8-11]. Second reason is that the intrinsic high frequencies of AFM dynamics [12]. It makes them distinct from ferromagnets. Due to these prominent properties, antiferromagnets have been used as magnetic recording media with good performance. Among various AFM materials the near equiatomic chemically ordered bcc-B2 (CsCl-type) bulk FeRh alloy is attracting intense interest due to its unusual first-order phase transition from AFM to FM order at≈350 K. This is accompanied with a volume expansion of ≈1% indicating a coupling between the spin and structural degrees of freedom. Together with the relativistic effects present in FeRh such as magnetocrystalline anisotropy, MCA, thermally assisted FeRh-based memory have been successfully fabricated. Most of density functional theory calculations to date have focused on the electronic structure properties solely of the bulk bcc structure under hydrostatic pressure [13]. On the other hand, FeRh thin films are grown expitaxially on MgO, BaTiO3, or piezoelectric substrates, where the lattice mismatch inevitably generates strain which may modulate the magnetic properties of the FeRh overlayer [14,15]. Recent ab initio electronic structure calculations have investigated the relative stability of the FM, "bcc-like-AFM", and "fcc-like-AFM" structures under epitaxial strain [16,17]. Therefore, it is of great importance to understand the effect of strain on the MCA of ultrathin FeRh thin films for further promoting their applications in AFM spintronics. In this work, we investigate the effect of epitaxial strain on the MCA of FeRh/MgO by performing ab initio density functional theory (DFT) electronic structure calculations. We find that the value of MCA and the direction of magnetic easy axis strongly depend on the strain leading to a spin re-orientation from an in- to out-of-plane magnetization orientation in the AFM phase and across the metamagnetic transition. The underlying mechanism is explained by analyzing the band structure and the strain-induced shifts of distribution of orbital characters. Our work demonstrates that strain engineering can serve as a viable and efficient approach in tuning the magnetization directions in FeRh.

## 2. Computational details

We use the Vienna ab initio simulation package (VASP) to perform the electronic structure calculations. The projector augmented wave formalism is adopted for describing the electron-ion interactions, and plane waves with a kinetic energy cutoff of 350 eV are used to expand the wave functions. The generalized gradient approximation (GGA) in the version of Perdew-Burke-Ernzerhof (PBE) is adopted for treating the electron exchange and correlation [18-22]. The GGA exchange correlation functional has been shown to provide a reasonable description of the structural properties of FeRh, in contrast to the local density approximation which yields incorrectly that the fcc-like AFM phase is the ground state. With a Monkhorst-Pack k-mesh of 31 x 31 x 31 for Brillouin zone (BZ) sampling, the calculated equilibrium lattice constants (a) of bulk G-AFM and FM phase FeRh are 2.995 Å and 3.012 Å, respectively, in good agreement with available experimental data. The ultrathin FeRh (001) film on the rock-salt MgO (001) substrate is modeled by a slab structure, as is shown in Fig. 1.

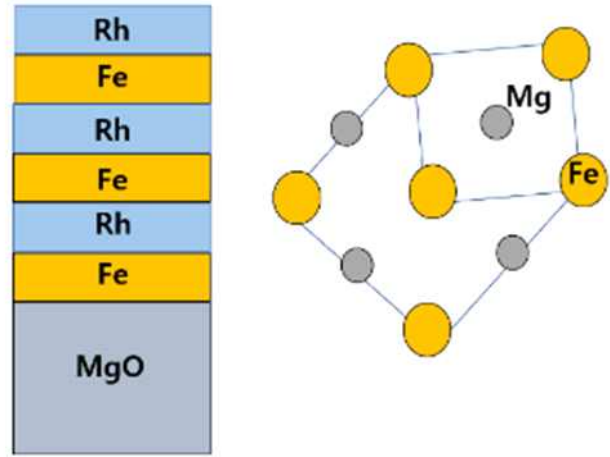

The slab supercell consists of five monolayers (ML) of FeRh with each ML consisting of two Fe or Rh atoms, which are placed on the top of a MgO slab. A 12 Å thick vacuum region is introduced in the supercell to avoid the artificial interactions between the slab and its images created by the periodic boundary conditions. The 110 axis of FeRh is aligned with the 100 axis of MgO and the O atoms at the FeRh/MgO interface are placed atop of the Fe atoms. In this work we consider the Fe-terminated interface and surface in both the G-AFM and FM phases, respectively. For each strain condition, the magnetic and electronic degrees of freedom and the atomic z-positions are fully relaxed until the maximum force acting on each single ion is less than 0.01 eV/Å. A Monkhorst-Pack k-mesh of 15 x 15 x 1 is used in the self-consistent DFT calculations followed by a 31 x 31 x 1 k-mesh for the scalar relativistic calculations with the spin-orbit coupling (SOC). The MCA per interfacial area is determined by [E[100]-E[001]] where E[100] (E[001]) is the total energy with SOC62included along the [100] and ([001]) directions, respectively.

## 3. Results and discussions

In Table I, we list the calculated values of the strain dependence of the spin magnetic moments ($\Delta m_s$), orbital magnetic moment differences ($m_o$) of the interfacial Fe atom ($Fe_i$) and surface Fe atom66($Fe_s$), respectively, for both the G-AFM and FM phases, respectively. We also list the MCA values and the total energy difference ($\Delta E$) between the the G-AFM and FM phases under different strain. Note because the two Fe atoms on each atomic plane in the G-AFM phase have opposite $m_s$.

TABLE I. Strain dependence of the spin magnetic moment (ms) and91orbital magnetic moment difference (mo = m [001] o m [100] o ) along the [001] and [100] directions, of92the interfacial ($Fe_i$) and surface ($Fe_s$) Fe atoms, for the AFM and FM phases, respectively. We list also93the strain dependence of the MCA values of the G-AFM and FM FeRh/MgO bilayer, as well as the94total energy difference between the two magnetic phases.

| strain(%) | G-AFM | | | | | FM | | | | | $\Delta E = E_{FM} - E_{AFM}$ |
|---|---|---|---|---|---|---|---|---|---|---|---|
| | $m_s(\mu_B)$ | | $\Delta m_o(10^{-2}\mu_B)$ | | MCA | $m_s(\mu_B)$ | | $\Delta m_o(10^{-2}\mu_B)$ | | MCA | |
| | $Fe_i$ | $Fe_s$ | $Fe_i$ | $Fe_s$ | ($erg/cm^2$) | $Fe_i$ | $Fe_s$ | $Fe_i$ | $Fe_s$ | ($erg/cm^2$) | (meV/Fe) |
| −0.5 | 3.028 | 3.152 | −3.6 | −5.3 | −0.47 | 3.068 | 3.180 | 5.9 | 2.3 | 0.45 | 19.3 |
| 0 | 3.027 | 3.153 | −2.7 | −5.7 | −0.07 | 3.069 | 3.184 | 5.8 | 1.9 | 0.55 | 20.3 |
| 0.5 | 3.026 | 3.153 | −1.7 | −5.7 | 0.44 | 3.076 | 3.186 | 5.2 | 1.6 | 0.30 | 22.0 |

we only list its magnitude. We find that for the range of strain considered here the G-AFM phase is more stable than the FM phase, where the energy difference $\Delta E = E_{FM} - E_{G-AFM}$= 20 meV/Fe regardless of the strain. Previous DFT calculations for free-standing FeRh films reported values of 25 meV/Fe, indicating that the MgO substrate slightly decreases the energy difference between the two phases. The $m_s$ values of $Fe_i$ and $Fe_s$ in the G-AFM and FM systems range from 3.02 to 3.19 $\mu_B$, which are close to the bulk value of 3.4 $\mu_B$, and depend weakly on strain. For the G-AFM phase the variation of MCA with strain is large and correlates well with the strain variation of $\Delta m_o$ of $Fe_i$ but not with that of $Fe_s$. For the G-AFM phase the compressive strain yields an in-plane magnetic easy axis while tensile strain induces an out-of-plane easy axis, where the spin reorientation occurs around 0. This strain induced MCA energy behavior is similar (opposite) to that in $CoFe_2O_4$ ($CoCr_2O_4$). Furthermore tensile strain induces an in-plane (out-of-plane) magnetic easy axis in Ta/FeCo/MgO (Au/FeCo/MgO) trilayers.

To understand the underlying mechanism of the strain effect, we plot in Fig. 2.

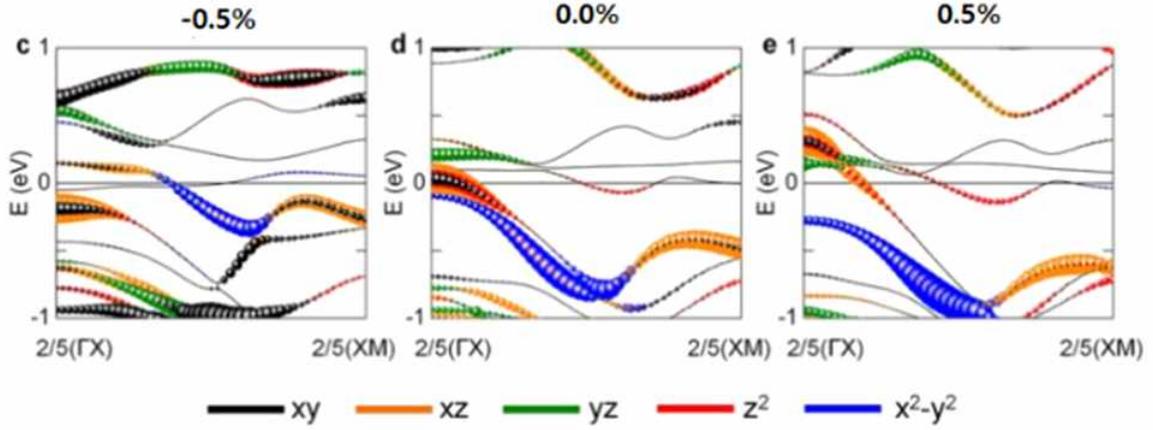

Figure 2 Energy- and k-resolved distribution of orbital character of the minority-spin bands of the interfacial Fe atom (Fei). k points with large negative or positive contributions to the total MCA are indicated by ΓX-XM (based on the second-order perturbation theory of MCA).

The energy and k-resolved distribution of the orbital character of the minority-spin bands of the spin-up interfacial iron $Fe_i$-derived $d_{xz,yz}$ and $d_{x2-y2}$ states for -0.50% strain. (The analysis of the spin-down $Fe_i$ atom is identical.) Within the second-order perturbation theory, the MCA can be expressed as:

$$\text{MCA} \propto \xi^2 \sum_{o,u} \frac{|\langle \Psi_o^\downarrow | \hat{L}_z | \Psi_u^\downarrow \rangle|^2 - |\langle \Psi_o^\downarrow | \hat{L}_{x/y} | \Psi_u^\downarrow \rangle|^2}{E_u^\downarrow - E_o^\downarrow} + \text{majority-spin term} + \text{spin-flip term}$$,

where the coupling matrix elemennt <·|·|·> denotes the SOC of the occupied state (o) with eigen energy Eo and the unoccupied state (u) with eigenenergy Eu through angular momentum operator $_{xy}$ or $_z$ (contributes positively) with proper selection rules. Since our analysis for the spin-up $Fe_i$ is in terms of the minority-spin bands, only the first term in Eq. (1) is important, because the majority spin bands are well below the Fermi energy. k points with large negative or positive contributions to the total MCA are indicated by ΓX-XM (based on the second-order perturbation theory of MCA). From Fig. 2 one

can see that around the ΓX and XM points the unoccupied $d_{yz,xz}$ states couple to the occupied $d_{x2-y2}$ state through $\ell_{xy}$, giving negative contributions to the total MCA. On the other hand, in a wide region around k between ΓX and XM the occupied and unoccupied $d_{yz,xz}$ states couple through $\ell_{z2}$ yielding large positive contributions. The sum over these k points gives the negative MCA of 0.47 erg/cm$^2$ for σ=0.50%. Note that around k points ΓX and XM cross the Fermi level and are prone to shift under different strain. In Fig. 3 we show the minority-spin projected density of states (PDOS) of the spin-up Fe$_i$ $d_{yz}$, $d_{xz}$, and $d_{x2-y2}$ orbitals for three strains. As one can see, with increasing there is a redistribution of PDOS on $d_{x2-y2}$ from unoccupied to occupied, indicating a down-shift of this orbital across the Fermi level around111k4. The PDOS distribution of $d_{xz}$ orbital in the range of [-0.5, 0.5] eV is slightly FIG. 3.

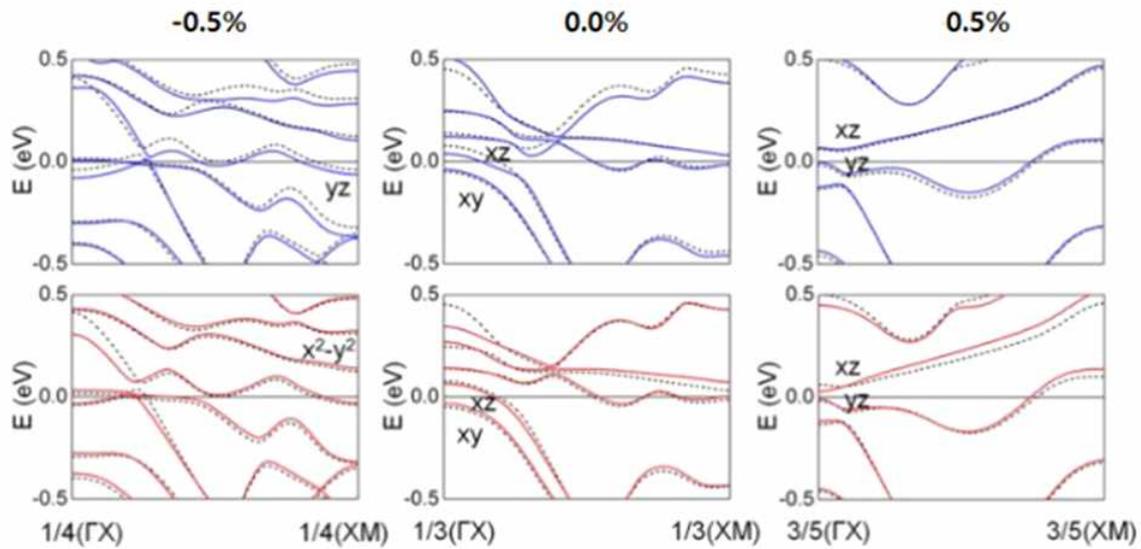

Projected density of states (PDOS) (minority) on the spin-up Fe$_i$ $d_{yz}$, $d_{xz}$, and $d_{x2-y2}$ orbitals for σ=-0.50, 0, and 0.50%, respectively. The PDOS of the $d_{xz}$ and $d_{x2-y2}$ are shifted by 1.5 and 3 units along the vertical axis, respectively. The peak of unoccupied PDOS shifts from 0.025 to 0.15 eV, suggesting a upward-shift of this orbital around ΓX and k between the ΓX and XM. The occupied PDOS distribution in the range of [0.1, 0.05] eV is largely enhanced due to the down-shift of the band. Overall, the band shifts will induce a large variation of MCA. Specifically, the band shifts around ΓX and XM decrease or disable the negative SOC while the band shifts at k3 enable the positive SOC. As a result, the MCA value changes from negative to positive when the strain changes from -0.5% to 0.5%.

## 4. Conclusion

We have studied the effect of epitaxial strain on the MCA of an ultrathin FeRh/MgO bilayer system by performing ab initio DFT electronic structure calculations. Under a compressive strain of -0.5%, the system possesses a large MCA value with the magnetic easy axis being in-plane. Detailed analyses based on the perturbation theory show this is due to the $<d_{yz,yz}|L_{yz,zx}|d_{x2-y2}>$ coupling matrix elements. When the strain changes from compressive to tensile, the induced band shifts and the concomitant redistributions of PDOS greatly change the contribution from each k point to the MCA, thus causing the reorientation of the magnetic easy axis from in- to out-of-plane. Our work demonstrates that strain engineering can be used to tailor the magnetic properties of AFM FeRh spintronic devices.